# Strategic Alliance for Blockchain Governance Game

SONG-KYOO KIM

## ABSTRACT

This paper deals with design of the alternative secure Blockchain network framework to prevent damages from an attacker. The concept of the strategic alliance of the management is applied on the top of the recent developed stochastic game framework. This new enhanced hybrid theoretical model has been developed based on the combination of the conventional game theory, the fluctuation theory and the *Blockchain Governance Game* to find best strategies towards preparation for preventing a network malfunction from an attacker by making the strategic alliance with other genuine miners. Analytically tractable results for decision making parameters are fully obtained which enable to predict the moment for operations and deliver the optimal number of the alliance with other nodes to protect the Blockchain network. This research helps for whom considers the initial coin offering or launching new blockchain based services with enhancing the security features by alliance with the trusted miners within the decentralized network.

# 1. INTRODUCTION

The Blockchain network is recently applied in a wide range of services and applications far beyond cryptocurrencies. Generally, the Blockchain is a distributed public digital ledger maintained by achieving the consensus among a number of nodes in a peer-to-peer network [1]. This decentralized peer-to-peer network has a strength of the blockchain which removes a number of security risks in the centralized network. Particularly, the verified transaction data is stored in a chain of blocks which is a basic data structure of the Blockchain, and the chain grows in an append-only manner with all new verified blocks [2]. The security of the Blockchain has been achieved through the distributed consensus of miners and this consensus is only reliable with the assumption that no single miner can hold more than 50 percent of the network computational power [2] and more than half miners are not aligned under single control. To gain its profit, the attacker either invests more than 50 percent of the computational power or governs more than half of the total nodes. If this assumption is broken, the distributed consensus are not valid any longer and it involves several operations including verifying transactions, disseminating blocks, and attaching blocks to the blockchain for maintaining this assumption. Although blockchain records are not unalterable, it could be considered as a secure network [3-4]. Recent researchers have improved the securities in the protocol levels and some researches have proposed the new protocol to prevent the 51 percent attack [5-7]. These protocol improvements might prevent the 51 percent attack but most solutions are limited because the implementations are robust by choosing the boundaries arbitrarily. Although the BGG is an innovative idea to protect the decentralized network from the majority attack, it requires to keep certain amount of genuine nodes for the proper protection of the network.

Because of this reason, alternative way to reserve genuine nodes which is called the *Strategic Alliance for Blockchain Governance Game* (SABGG) has been designed instead of keeping honest nodes by the defender. In this paper, the *Strategic Alliance for BGG* is newly proposed. In general, the strategic alliance is an agreement within two or more parties to pursue a set of agreed upon objectives needed and it has emerged to solve many company business problems [11]. The concept of the strategic alliance has been adapted to improve the BGG based network security. Atypical case that an attacker to trying to build an alternative chains faster than regular miners is considered [10]. Instead of keeping genuine nodes, the defender (or controller) makes the agreement with other genuine miners as a strategic alliance member and the defender requests for taking the alliance action prior an attack happens. The genuine miners could either accept the request or reject it when other miners receive the alliance action request. The normal nodes within the strategic alliance members become reserved genuine nodes for backups if they agrees the alliance at the moment of the request. It is noted the total number of alliance member may not same as the actual honest nodes for backup in the BGG. Each request moment of the alliance action may have different members to accept each request.

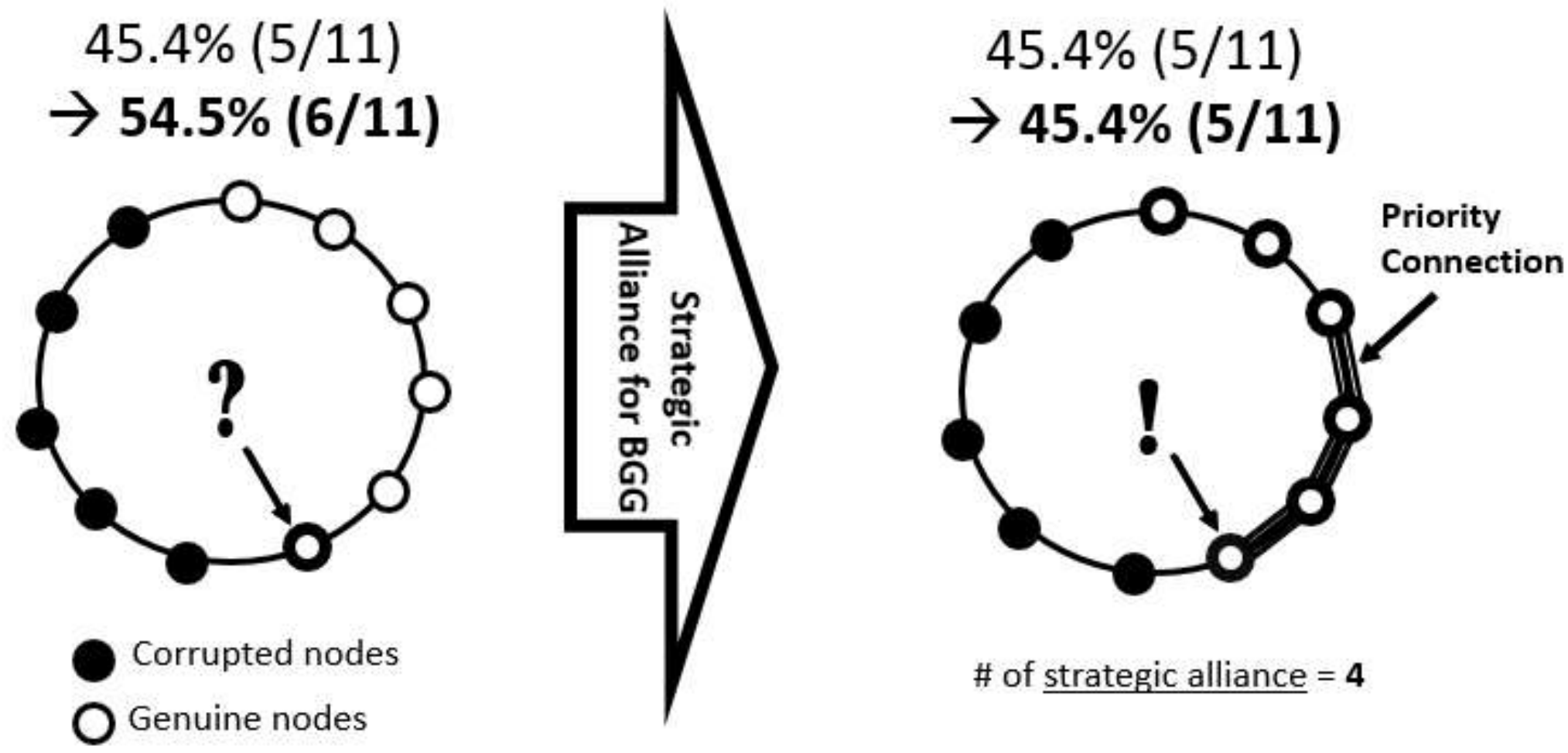

**Fig. 1.** Strategic Alliance for Blockchain Governance Game

The new results of the SABGG are given as joint functionals between two players of the predicted time of the first observed threshold which is crossing the half of the total nodes (i.e., 51 percent) along with values of each component upon this time [10].

The paper is organized as follows: In Section 2, a basic model is introduced and the key technique is preserving the wide generality of our result. Once a corrupted blocks are generated, the model predicts how many blocks will be generated and the moment until more than a half nodes are corrupted by an attacker. In other words, the paper have modeled processes of cumulated corrupted blocks of nodes and weights to predict both time and size of the nodes which governed by an attacker. The technique delivers the results under a composition of the operator: R-operator which has been introduced in some past work of Dshalalow [12-13]. Basically, the initial setup is same in the BGG [10] except for the security policy. The moment of the decision making is also analyzed in Section 2. Unlike the BGG model, the SABGG improves the Blockchain security by allying genuine miners instead of keeping honest nodes by the defender. The framework for setting up the mixed strategic game is provided in Section 3. The optimal values and the cost of the strategic alliance for the blockchain governance game are analytically calculated. Lastly, the special case of the SABGG is introduced in Section 4. The observation process could have the memoryless properties which implies that a defender (i.e., a blockchain based service provider) does not spend additional cost of keeping the past information. This implication also helps the programming implementation of a blockchain based service by reducing the analysis complexities.

## 2. Stochastic Model For Blockchain Network

### 2.1 Basic Stochastic Model

As it mentioned in the previous section, the setup of the basic model is similar as the BGG model [10]. The antagonistic game of two players (called "Corrupted" and "Genuine") is provided and players compete to build the blocks either for genuine (or honest) or corrupted (or false) ones.

$$\mathcal{C} := \sum_{j\geq 0} J_j \varepsilon_{u_j},\ u_0(=0) < u_1 < u_2 < \cdots,\ \text{a.s.} \tag{2.1}$$

$$\mathcal{G} := \sum_{k\geq 0} K_k \varepsilon_{v_k},\ v_0(=0) < v_1 < v_2 < \cdots,\ \text{a.s.} \tag{2.2}$$

are $\mathcal{F}_C$-measurable and $\mathcal{F}_G$-measurable marked Poisson processes with respective intensities $\lambda_c$ and $\lambda_g$ with the point independent marking. It is assumed that a third-party observation point process is equivalent with the duration of PoW (Proof-of-Work) or PoS (Proof-of-Sake) completions before specifying (2.1) and (2.2) :

$$\mathcal{U} := \sum_{i\geq 0} \varepsilon_{t_i},\ t_0(>0)), t_1, \ldots. \tag{2.3}$$

Player C builds the blocks with false transactions (e.g., double spending) and Player G generate the blocks which contain the correct transactions. Both players races to build their blocks (either genuine or corrupted). The processes $\mathcal{C}$ and $\mathcal{G}$ are specified by their transforms:

$$\mathbb{E}\big[y^{\mathcal{C}(u)}\big] = e^{\lambda_c(u)(y-1)},\ \mathbb{E}\big[z^{\mathcal{G}(v)}\big] = e^{\lambda_g(v)(z-1)}. \tag{2.4}$$

and

$$(C(t), G(t)) := \mathcal{C} \otimes \mathcal{G}([0, t_i]),\ i = 0, 1, \ldots, \tag{2.5}$$

forms an observation process upon $\mathcal{C} \otimes \mathcal{G}$ embedded over $t$, with respective increments

$$(J_i, K_i) := \mathcal{C} \otimes \mathcal{G}([t_{i-1}, t_i]),\ i = 1, 2, \ldots, \tag{2.6}$$

and

$$J_0 = C_0,\ K_0 = G_0. \tag{2.7}$$

The observation process is formalized as

$$\mathcal{C}_t \otimes \mathcal{G}_t := \sum_{i\geq 0} (J_i, K_i) \varepsilon_{t_i}, \tag{2.8}$$

where

$$\mathcal{C}_t = \sum_{i\geq 0} J_i \varepsilon_{t_i},\ \mathcal{G}_t = \sum_{i\geq 0} K_i \varepsilon_{t_i}, \tag{2.9}$$

and it is with position dependent marking and with $J_i$ and $K_i$ being dependent with the notation

$$U_i := t_i - t_{i-1},\ i = 0, 1, \ldots,\ t_{-1} = 0, \tag{2.10}$$

and

$$\alpha(y, z) = \mathbb{E}\big[y^{J_i} \cdot z^{K_i}\big],\ y > 0,\ z > 0. \tag{2.11}$$

By using the double expectation [13],

$$\alpha(y, z) = \alpha(\lambda_c(1-y) + \lambda_g(1-z)) \tag{2.12}$$

and

$$\alpha_0(y,z) = \mathbb{E}\left[y^{C_0} z^{G_0}\right] = \alpha_0(\lambda_c(1-y) + \lambda_g(1-z)) \tag{2.13}$$

where

$$\alpha(\theta) = \mathbb{E}\left[e^{-\theta U_1}\right],\ \alpha_0(\theta) = \mathbb{E}\left[e^{-\theta t_0}\right] \tag{2.14}$$

To further formalize the game, the *exit indexes* are defined as follows:

$$\nu_1 := \mathit{inf}\left\{k : C_k\ (= C_0 + J_1 + \cdots + J_k) \geq \left(\tfrac{N}{2} - \eta\right)\right\}, \tag{2.15}$$

$$\nu := \mathit{inf}\left\{j : C_j\ (= C_0 + J_1 + \cdots + J_j) \geq \left(\tfrac{N}{2}\right)\right\}, \tag{2.16}$$

$$\nu_2 := \mathit{inf}\left\{j : C_j\ (= C_0 + J_1 + \cdots + J_j) - B_\eta \geq \left(\tfrac{N}{2}\right)\right\}, \tag{2.17}$$

$$\mu_1 := \mathit{inf}\left\{i : G_i (= G_0 + K_1 + \cdots + K_i) + B_\eta \geq \left(\tfrac{N}{2}\right)\right\}, \tag{2.18}$$

$$\mu := \mathit{inf}\left\{l : G_l (= G_0 + K_1 + \cdots + G_l) \geq \left(\tfrac{N}{2}\right)\right\}. \tag{2.19}$$

The game is over when player C beats the game at time $t_\nu$ without the strategic alliance with other nodes (i.e., the action of security is executed). With the strategic alliance, player C beats the game at time $t_\nu$ unless player G beats at $t_\mu$ take place earlier. Thus game is over at $\mathit{min}\{\nu_1, \nu, \nu_2, \mu_1, \mu\}$. However, the game with player C beating first without the security action. The $\sigma$-subalgebra of the process $(\mathcal{C}, \mathcal{G})$ can be analogously as $\mathcal{F}(\Omega) \cap \{(\nu_1) < \nu < \nu_2 ( < \mu_1) < \mu\}$. We shall be targeting the confined game in the view point of player C. The first passage time $t_\nu$ is the associated exit time from the confined game and the formula (2.8) will be modified as

$$\overline{\mathcal{C}_t} \otimes \overline{\mathcal{G}_t} := \sum_{n \geq 0}^{\nu} (J_n, K_n) \varepsilon_{t_n} \tag{2.20}$$

which gives an exact definition of the model observed until $t_\nu$ without the strategic alliance action. The joint functional of the blockchain network model is as follows:

$$\begin{aligned} \Theta_{\frac{N}{2}} &= \Theta_{\{\nu,\nu_1,\mu\}}(\zeta, y_0, y_1, b, z_0, z_1) \\ &= \mathbb{E}\left[\zeta^{\nu} \cdot y_0^{C_{\nu-1}} \cdot y_1^{C_\nu} \cdot b^{C_\nu - B_\eta} \cdot z_0^{G_{\mu-1}} \cdot z_1^{G_\mu} \mathbf{1}_{\{\nu < \nu_2 < \mu\}}\right] \end{aligned} \tag{2.21}$$

where $N$ indicates the total number of nodes (or ledgers) in the Blockchain network. This functional will represent the status of attackers and honest nodes upon the exit time $t_\nu$. The latter is of particular interest, we are interested in not only the prediction of catching up the blocks by attackers but also one observation prior to this. The Theorem BGG-2 establishes an explicit formula for $\Theta_{\frac{N}{2}}$ with (2.11) and (2.13). The first exceed model by Dshahalow and the operator is defined as follows:

$$\begin{aligned} &\mathcal{R}_{(a,b,c)}\left[g(a,b,c)\right](q,r,s) \\ &\qquad := (1-q)(1-r)(1-s)\left\{\sum_{a\geq 0}\sum_{b\geq 0}\sum_{c\geq 0} g(a,b,c) q^a r^b s^c\right\}, \end{aligned} \tag{2.22}$$

then

$$g(a,b,c) = \Re_{(q,r,s)}\left[\mathcal{R}_{(a,b,c)}\left\{g(a,b,c)\right\}(q,r,s)\right](a,b,c) \tag{2.23}$$

where $\{g(a,b,c)\}$ is a sequence, with the inverse

$$\mathfrak{R}_{(q,r,s)}^{(a,b,c)}(\bullet) = \begin{cases} \left(\frac{1}{a!\cdot b!\cdot c!}\right) \lim\limits_{(q,r,s)\to 0} \frac{\partial^a \partial^b \partial^c}{\partial q^a \partial r^b \partial s^c} \frac{1}{(1-q)(1-r)(1-s)}(\bullet), & a,b,c \geq 0 \\ 0, & \textit{otherwise}. \end{cases} \tag{2.24}$$

**Theorem 1 (BGG-2):** the functional $\Theta_{\frac{N}{2}}$ of the process of (2.21) satisfies following expression:

$$\Theta_{\lceil\frac{N}{2}\rceil} = \mathfrak{R}_{(q,r,s)}\, \Lambda\left(\left\lceil\frac{N}{2}\right\rceil, \left\lceil\frac{N}{2}\right\rceil, \left\lceil\frac{N}{2}\right\rceil\right), \tag{2.25}$$

because

$$\Lambda = \Lambda(q,r,s) = \mathcal{R}_{(q,r,s)}[\Theta_n](q,r,s), \tag{2.26}$$

and

$$\Lambda = \left\{\sigma_\eta \cdot \beta\left[\frac{1-\beta^1}{1-\beta}\right]\left[\alpha_0^1 - \alpha_0 + \frac{\zeta\alpha_0}{1-\zeta\alpha}\left(\alpha^1 - \alpha\right)\right]\right\}. \tag{2.27}$$

From (2.23) and (2.44), we can find the PGFs (probability generating functions) of $C_{\nu-1}$ (and $C_\nu$) and the *exit index*:

$$\mathbb{E}[\zeta^\nu] = \Theta_{\lceil\frac{N}{2}\rceil}(\zeta,1,1,1,1,1). \tag{2.46}$$

$$\mathbb{E}\left[y_0^{C_{\nu-1}}\right] = \Theta_{\lceil\frac{N}{2}\rceil}(1,y_0,1,1,1,1), \tag{2.47}$$

$$\mathbb{E}\left[y_1^{C_\nu}\right] = \Theta_{\lceil\frac{N}{2}\rceil}(1,1,y_1,1,1,1), \tag{2.48}$$

$$\mathbb{E}\left[b^{C_\nu - B_\eta}\right] = \Theta_{\lceil\frac{N}{2}\rceil}(1,1,1,b,1,1), \tag{2.49}$$

The moment of making a decision $\tau_{\nu-1}$ could be found from (2.4), (2.10) and (2.46):

$$\mathbb{E}[\nu] = \frac{\partial}{\partial\zeta}\Theta_{\lceil\frac{N}{2}\rceil}(\zeta,1,1,1,1,1)\Big|_{\zeta=1}, \tag{2.50}$$

$$\mathbb{E}[t_{\nu-1}] = \mathbb{E}[t_0] + \mathbb{E}[U_1](\mathbb{E}[\nu] - 1). \tag{2.51}$$

## 3. STRATEGIC ALLIANCE FOR BLOCKCHAIN GOVERNANCE GAME

### 3.1 Preliminaries

A two-person mixed strategy game is considered, and the player G (i.e., a defender who keeps genuine nodes) is the person who has two strategies at the observation moment, one step before attackers complete to generate alternative chains with corrupted transactions. Two strategies, (1) *DoNothing* – doing nothing, which implicates that the Blockchain networks are running as usual, and (2) *Action* – taking the preliminary action for avoiding attacks by adding genuine nodes, for player G are considered. The responses of player C would be either "Not burst" or "Burst." The cost for the strategic alliance is $c(\eta)$ and the network bursts if the attacks succeed to generate alternative blocks. The whole value of the tokens (or coins) $V$ will be lost and this value might be equivalent with the value by ICO (Initial Coin Offering). It still has the chance to burst even if the defender (or the provider) adds the genuine nodes by the agreement with the strategic partners. In this case, the cost will be not only the token value but also the cost for the strategic alliance and the normal form of games is as follows:

$$\begin{aligned} &\text{. Players:} \quad \boldsymbol{M} = \{c,\ g\}, \\ &\text{. Strategy sets:} \\ &\qquad \boldsymbol{s_c} = \{\text{"}\textit{NotBurst}\text{"}, \text{"}\textit{Burst}\text{"}\}, \\ &\qquad \boldsymbol{s_g} = \{\text{"}\textit{DoNothing}\text{"}, \text{"}\textit{Action}\text{"}\}, \end{aligned} \tag{3.1}$$

Based on the above conditions, the general cost matrix at the prior time to be burst $t_{\nu-1}$ could be composed as follows:

| | *NotBurst* $(1-q(s_g))$ | *Burst* $(q(s_g))$ |
|---|---|---|
| *DoNothing* | $0$ | $V$ |
| *Action* | $c(\eta)$ | $c(\eta)+V$ |

**Table 1.** Cost matrix

where $q(s_g)$ is the probability of bursting blockchain network (i.e., an attacker wins the game) and it depends on the strategic decision of player G (a defender):

$$q(s_g) = \begin{cases} \mathbb{E}\left[\mathbf{1}_{\{C_\nu \geq \frac{N}{2}\}}\right], & s_g = \{\textit{DoNothing}\}, \\ \mathbb{E}\left[\mathbf{1}_{\{C_\nu - B_\eta \geq \frac{N}{2}\}}\right], & s_g = \{\textit{Action}\}. \end{cases} \tag{3.2}$$

It is noted that the cost for the alliance (i.e., "Action" strategy of player G) should be better than the other strategy. Otherwise, player G does not have to spend the cost of the governance. The total number of the alliance nodes $\eta$ is fixed but the agreement $B$ is uncertain when the alliance is actually requested. Let us consider the average acceptance rate within the strategic alliance $\varrho$ then the probability of the actual agreement $B$ is the binomial distribution and $\sigma_\eta$ from (2.43) is determined as follows:

$$\sigma_\eta = \mathbb{E}\left[b^{-B}\right] = \left(\frac{\varrho}{b} - (1-\varrho)\right)^\eta. \tag{3.3}$$

The total number of nodes for the strategic alliance $\eta$ depends on the cost function and the optimal number of alliance members $\eta^*$ could be found as follows:

$$\eta^* = \textit{inf}\{\eta \geq 0 : \mathfrak{S}_{\textit{NoA}}\left(q^0\right) \geq \mathfrak{S}_{\textit{Act}}(\eta)\}, \tag{3.4}$$

where (at the moment $t_{\nu-1}$),

$$\mathfrak{S}_{\textit{NoA}}\left(q^0\right) = V \cdot q^0, \tag{3.5}$$

$$\mathfrak{S}_{\textit{Act}}(\omega) = c_\eta\left(1 - q_\eta^1\right) + (c_\eta + V)q_\eta^1, \tag{3.6}$$

$$q^0 = \mathbb{E}\left[\mathbf{1}_{\{C_\nu \geq \frac{N}{2}\}}\right],\ q_\eta^1 = \mathbb{E}\left[\mathbf{1}_{\{C_\nu - B_\eta \geq \frac{N}{2}\}}\right], \tag{3.7}$$

where, from (2.48),

$$P\{C_\nu - B_\eta = k\} = \begin{cases} \lim\limits_{b\to 0} \frac{1}{k!}\frac{\partial^k}{\partial b^k}\Theta_{\lceil \frac{N}{2}\rceil}(1,1,1,b,1,1), & k = 0,\dots,N-1, \\ 1 - \sum\limits_{k=0}^{N-1} P\{C_\nu - B_\eta = k\}, & k = N. \end{cases} \tag{3.8}$$

### 3.2. Strategic Alliance For Blockchain Governance Game

We would like to design the enhanced blockchain network governance that can take the action at the decision making moment $t_{\nu-1}$. The governance in the blockchain is followed by the decision making parameter. It also means that we will not take any action until the time $t_{\nu-1}$ and it still have the chance that all nodes are governed by an attacker if the attacker catches more than the half of nodes at $t_{\nu-1}$ (i.e., $\{C_{\nu-1} \geq \frac{N}{2}\}$. If the attacker catches the less than half of all nodes at $t_{\nu-1}$ (i.e., $\{C_{\nu-1} < \frac{N}{2}\}$, then the defender could take the action to avoid the attack at $t_{\nu-1}$. The total cost for developing the enhanced blockchain network is as follows:

$$\begin{aligned} \mathfrak{S}(q^0,\eta)_{Total} &= \mathbb{E}\Big[\mathfrak{S}_{Act}(\eta)\cdot \mathbf{1}_{\{C_{\nu-1}<\frac{N}{2}\}} + \mathfrak{S}_{NoA}(q^0)\cdot \mathbf{1}_{\{C_{\nu-1}\geq\frac{N}{2}\}}\Big] \\ &= \big(c_\eta(1-q_\eta^1) + (c_\alpha + V)q_\eta^1\big)p_{c_{-1}} + V\cdot q^0(1-p_{c_{-1}}), \end{aligned} \tag{3.9}$$

where

$$p_{c_{-1}} = \boldsymbol{P}\Big\{C_{\nu-1} < \frac{N}{2}\Big\} = \sum_{k=0}^{\lfloor \frac{N}{2}\rfloor}\boldsymbol{P}\{C_{\nu-1} = k\}. \tag{3.10}$$

Because $\Theta_{\lceil \frac{N}{2}\rceil}(1,y_0,1,1,1,1)$ from (2.21) is the probability generating function of $C_{\nu-1}$, the probability mass could be found as follows:

$$\boldsymbol{P}\{C_{\nu-1} = k\} = \lim_{y_0\to 0}\frac{1}{k!}\frac{\partial^k}{\partial y_0^k}\Theta_{\lceil \frac{N}{2}\rceil}(1,y_0,1,1,1,1),\ k = 0,\dots,\left\lfloor \frac{N}{2}\right\rfloor. \tag{3.11}$$

## 4. SPECIAL CASE: MEMORYLESS OBSERVATION PROCESS

It is assumed that the observation process has the memoryless properties which might be a special condition but very practical for actual implementation on the *Strategic Alliance for BGG*. It implies that the defender (or a service provider) does not spend additional cost of storing the past information. To build the cost function of the blockchain governance, we can find explicit solutions of $q^0$, $p_{c_{-1}}$ and the moment of the decision making after finding the first exceed level index $\mathbb{E}[\nu]$, the probability (generating function) of the number of blocks at the moment $t_\nu$ $\left(\mathbb{E}\left[y_1^{C_\nu}\right]\right)$ and $\tau_{\nu-1}$ $\left(\mathbb{E}\left[y_0^{C_{\nu-1}}\right]\right)$. Recall from (2.22), the operator is defined as follows:

$$G(q) = \mathcal{R}_a\left[f(a)\right](q) \coloneqq (1-u)\sum_{a\geq 0} f(a)q^a, \tag{2.22}$$

$$\begin{aligned}
&\mathcal{R}_{(a,b,c)}\left[f_1(a)f_2(b)f_3(c)\right](q,r,s) \\
&\quad \coloneqq (1-q)(1-r)(1-s)\sum_{a\geq 0}\sum_{b\geq 0}\sum_{c\geq 0} f_1(a)f_2(b)f_3(c)q^a r^b s^c \\
&\quad = \mathcal{R}_a[f_1(a)]\mathcal{R}_b[f_2(b)]\mathcal{R}_c[f_3(c)],
\end{aligned} \tag{4.1}$$

then

$$f(a,b,c) = \mathfrak{R}^{(a,b,c)}_{(q,r,s)}\left[\mathcal{R}_{(a,b,c)}\left\{f(a,b,c)\right\}\right], \tag{4.2}$$

$$f_1(a)f_2(b)f_3(c) = \mathfrak{R}^a_q[\mathcal{R}_a\left\{f_1(a)\right\}]\mathfrak{R}^b_r[\mathcal{R}_b\left\{f_2(b)\right\}]\mathfrak{R}^c_s[\mathcal{R}_c\left\{f_3(c)\right\}], \tag{4.3}$$

where $\{f(x), (f_1(x)f_2(y)f_3(z)),\}$ are a sequence, with the inverse (4.3) and

$$\mathfrak{R}^a_q(\,\bullet\,) = \begin{cases} \frac{1}{a!}\lim\limits_{q\to 0}\frac{\partial^a}{\partial q^a}\frac{1}{(1-q)}(\,\bullet\,), & a\geq 0, \\ 0, & \textit{otherwise,} \end{cases} \tag{4.4}$$

and

$$\mathfrak{R}^{(a,b,c)}_{(q,r,s)}[G_1(q)G_2(r)G_3(s)] = \mathfrak{R}^a_q[G_1(q)]\mathfrak{R}^b_r[G_2(r)]\mathfrak{R}^c_s[G_3(s)]. \tag{4.5}$$

It is also noted that the formulas (2.11)-(2.14) could be rewritten as follows:

$$\alpha(y,z) = \alpha(\lambda_c(1-y)+\lambda_g(1-z)) = \alpha_c(y)\cdot\alpha_g(z), \tag{4.7}$$
$$\alpha_c(y) = \alpha(\lambda_c(1-y)), \tag{4.8}$$
$$\alpha_g(z) = \alpha(\lambda_g(1-z)), \tag{4.9}$$

and

$$\alpha_0(y,z) = \alpha_0(\lambda_c(1-y)+\lambda_g(1-z)) = \alpha^0_c(y)\cdot\alpha^0_g(z), \tag{4.10}$$
$$\alpha^0_c(y) = \mathbb{E}\left[y^{C_0}\right] = \alpha_0(\lambda_c(1-y)), \tag{4.11}$$
$$\alpha^0_g(z) = \mathbb{E}\left[z^{G_0}\right] = \alpha_0(\lambda_g(1-z)), \tag{4.12}$$

from (2.35)-(2.43),

$$\Phi = \Phi_c\cdot\Phi_g \coloneqq \alpha_c(y_0y_1bqr)\cdot\alpha_g(z_0z_1s), \tag{4.13}$$
$$\Phi_0 = \Phi^0_c\cdot\Phi^0_g \coloneqq \alpha_0(y_0y_1bqr)\cdot\alpha_0(z_0z_1s), \tag{4.14}$$

$$\alpha = \alpha_c\alpha_g \coloneqq \alpha_c(y_1bq)\alpha_g(z_1), \tag{4.15}$$
$$\alpha_0 = \alpha^0_c\cdot\alpha^0_g \coloneqq \alpha^0_c(y_1bq)\alpha^0_g(z_1), \tag{4.16}$$

$$\alpha^1 \coloneqq \alpha(y_1b,z_1) = \alpha_c(y_1b)\alpha_g(z_1), \tag{4.17}$$
$$\alpha^1_0 \coloneqq \alpha_0(y_1b,z_1) = \alpha^0_c(y_1b)\alpha^0_g(z_1), \tag{4.18}$$

$$\beta \coloneqq \alpha(br,s) = \alpha_c(br)\alpha_g(s), \tag{4.19}$$
$$\beta^1 \coloneqq \alpha(r,1) = \alpha_c(r). \tag{4.20}$$

Since, the observation process has the memoryless properties, the process is exponentially distributed and the functionals from (4.7)-(4.12) are as follows:

$$\alpha_c^0(q) = \frac{1}{(1+\widetilde{\alpha_0}\cdot\lambda_c)-\widetilde{\alpha_0}\cdot\lambda_c q} = \frac{b_c^0}{1-a_c^0\cdot q}, \tag{4.21}$$

$$\alpha_c(q) = \frac{1}{(1+\widetilde{\alpha}\cdot\lambda_c)-\widetilde{\alpha}\cdot\lambda_c q} = \frac{b_c}{1-a_c\cdot q}, \tag{4.22}$$

$$\alpha_g^0(s) = \frac{1}{(1+\widetilde{\alpha_0}\cdot\lambda_g)-\widetilde{\alpha_0}\cdot\lambda_g s} = \frac{b_g^0}{1-a_g^0\cdot s}, \tag{4.23}$$

$$\alpha_g(s) = \frac{1}{(1+\widetilde{\alpha}\cdot\lambda_g)-\widetilde{\alpha}\cdot\lambda_g s} = \frac{b_g}{1-a_g\cdot s}, \tag{4.24}$$

$$b_c^0 = \frac{1}{(1+\widetilde{\alpha_0}\cdot\lambda_c)},\ a_c^0 = \frac{\widetilde{\alpha_0}\cdot\lambda_c}{(1+\widetilde{\alpha_0}\cdot\lambda_c)}, \tag{4.25}$$

$$b_c = \frac{1}{(1+\widetilde{\alpha}\cdot\lambda_c)},\ a_c^0 = \frac{\widetilde{\alpha}\cdot\lambda_c}{(1+\widetilde{\alpha}\cdot\lambda_c)}, \tag{4.26}$$

$$b_g^0 = \frac{1}{(1+\widetilde{\alpha_0}\cdot\lambda_g)},\ a_c^0 = \frac{\widetilde{\alpha_0}\cdot\lambda_g}{(1+\widetilde{\alpha_0}\cdot\lambda_g)}, \tag{4.27}$$

$$b_g = \frac{1}{(1+\widetilde{\alpha}\cdot\lambda_g)},\ a_c^0 = \frac{\widetilde{\alpha}\cdot\lambda_g}{(1+\widetilde{\alpha}\cdot\lambda_g)}, \tag{4.28}$$

where

$$\widetilde{\alpha_0} = \mathbb{E}[t_0],\ \widetilde{\alpha} = \mathbb{E}[U_k]. \tag{4.29}$$

### 4.1. The Probability Generating Function of $\nu$

The *exit index* (aka, the *first exceed level index*) is the most important factor to be fully analyzed because the decision factors including the marginal mean of $t_{\nu-1}$, $C_\nu$ and $C_{\nu-1}$ could be calculated easily if the exit index is fully analyzed. It could be straight forward once the *exit index* is found from (2.29)-(2.34), (2.36) and (3.15)-(3.26):

$$\mathbb{E}[\zeta^\nu] = \Theta_{\lceil\frac{N}{2}\rceil}(\zeta,1,1,1,1,1) = R^1 + R^2 - R^3, \tag{4.30}$$

where

$$R^1 = \left\{\frac{a_g b_c b_g}{1-b_c b_g}\right\}\left\{\mathit{\Xi}_{\frac{N}{2}}(0)\right\}\left(1 - b_c + b_c\left(\sum_{l\geq 0}^{\frac{N}{2}}(a_c^0)^l\right)\right) - \left\{\mathit{\Xi}_{\frac{N}{2}}(0)\right\}\left(\frac{b_c^0}{a_g}\right)\left\{\sum_{k\geq 0}^{\frac{N}{2}}\left(1+\left(\frac{a_g b_c b_g}{1-b_c b_g}\right)\right)^{k+1}\right\}\left(\sum_{l\geq 0}^{\frac{N}{2}}(a_c^0)^l\right), \tag{4.31}$$

$$R_{22} = \left(\frac{\zeta b_c^0 b_c b_g^0}{a_c^0\{1-b_c b_g - a_c\}-a_g(a_c+1)}\right) \cdot \sum_{l\geq 0}\left[\left(a_g^0\right)^l\left\{\Xi^{\frac{N}{2}-l}(\zeta) - (a_g)\Xi^{\frac{N}{2}-l-1}(\zeta)\right\}\left\{\sum_{j=l}\binom{j}{l}\left(\frac{1}{a_g^0}\right)^{j+1}\right\}\right], \tag{4.32}$$

$$R^3 = \left\{\frac{\zeta b_c (b_c^0)^2}{\{1-b_c b_g - a_c\}-a_g(a_c+1)}\right\} \cdot \sum_{h\geq 0}\binom{k}{h}\left(a_g^0\right)^{h-1}\left\{\sum_{k=h}\left(\frac{a_c^0}{a_g^0}\right)^k\right\}\left[\Xi^{\frac{N}{2}-h}(\zeta) - (a_g)\Xi^{\frac{N}{2}-h-1}(\zeta)\right] \tag{4.33}$$

$$
+\left(\frac{\zeta b_c^0 b_g^0 (b_c)^2}{a_c^0(1-a_c)}\right)\left(\frac{1}{(1-b_c b_g)-a_c-a_g(1-a_c)}\right)
$$
$$
\cdot\sum_{h\geq 0}\binom{k}{h}\left(a_g^0\right)^h\sum_{k\geq h}\left(\frac{a_c^0}{a_g^0}\right)^{k+1}\left[\Xi^{\frac{N}{2}-h}(0)-\left(a_g\right)^k\Xi^{\frac{N}{2}-h-1}(\zeta)\right],
$$

and

$$
\Xi_j(0)=\ \Xi_{\frac{N}{2}}(0)=\left\{\sum_{u=0}^{m}\left(\frac{\left(\frac{N}{2}\right)!}{\left(\left(\frac{N}{2}\right)-u\right)!}\right)\prod_{h=1}^{\left(\frac{N}{2}\right)}\left(\frac{h!}{1-\left(\frac{a_c}{1-b_c b_g}\right)}\right)\right\}, \tag{4.34}
$$

$$
\Xi^m(\zeta)\coloneqq\Xi^m(\zeta)=\left\{\sum_{u=0}^{m}\left(\frac{m!}{(m-u)!}\right)\ \prod_{l=1}^{m}\left(\frac{l!}{1-\left(\frac{a_g^0}{1-\zeta b_c^0 b_g^0}\right)}\right)\right\}. \tag{4.35}
$$

In addition, the proof of the calculations from (4.31)-(4.35) could found in Appendix A.

**4.2. The Marginal Means of The Decision Making Parameters**

As it is mentioned above, atypical decision making parameters are $\nu,\ t_{\nu-1},\ C_\nu$ and $C_{\nu-1}$. Although all decision making parameters could be fully analyzed, using the marginal mean of the parameters is occasionally more efficient than finding an explicit PGF of each parameter. The marginal mean of the decision making parameters could be found as follows:

$$
\mathbb{E}[\nu]=\frac{\partial}{\partial\zeta}\Theta_{\lceil\frac{N}{2}\rceil}(\zeta,1,1,1,1{,}1)\Big|_{\zeta=1}, \tag{2.49}
$$

$$
\mathbb{E}[t_{\nu-1}]=\mathbb{E}[t_0]+\mathbb{E}[U_1](\mathbb{E}[\nu]-1), \tag{2.50}
$$

$$
\mathbb{E}[C_\nu]=\mathbb{E}[\mathbb{E}[C_\nu|\nu]]=\mathbb{E}[C_0]+\mathbb{E}[\nu-1]\mathbb{E}[J_k], \tag{4.36}
$$

$$
\mathbb{E}[C_\nu-B_\eta]=\mathbb{E}[C_\nu]-\mathbb{E}[B_\eta], \tag{4.37}
$$

$$
\mathbb{E}[C_{\nu-1}]=\mathbb{E}[\mathbb{E}[C_\nu|\nu-1]]=\mathbb{E}[C_0]+\mathbb{E}[\nu-2]\mathbb{E}[J_k]. \tag{4.38}
$$

Recall from (3.2), the probability of bursting blockchain network (i.e., an attacker wins the game) under the memoryless properties becomes the Poisson Compound process:

$$
q(s_g)=\begin{cases}\sum\limits_{k>\frac{N}{2}}\mathbb{E}\left[\mathbf{1}_{\{C_\nu=k\}}\right], & s_g=\{\mathit{DoNothing}\},\\[2ex] \mathbb{E}\left[\mathbb{E}\left[\sum\limits_{k>\frac{N}{2}+B_\eta}\mathbb{E}\left[\mathbf{1}_{\{C_\nu=k\}}\right]\middle|B_\eta\right]\right], & s_g=\{\mathit{Action}\},\end{cases} \tag{4.39}
$$

where

$$\mathbb{E}\big[\mathbf{1}_{\{C_\nu=k\}}\big] = \mathbb{E}\left[\mathbb{E}\left[\frac{\lambda_c t_\nu}{k!}\cdot e^{-\lambda_c t_\nu}\middle| t_\nu\right]\right] \tag{4.40}$$

### 4.2. Linear Programming Practice

A network security in a semipublic blockchain based service with the token offering is considered in this subsection. Although the Blockchain network is designed to be decentralized, the service provider should have some strategy to avoid attacks not only from outsiders but also from insiders. As it is mentioned, the strategy for managing the network reliability is the strategic alliance with other trusted miners to give the less chance that an attacker catches blocks with corrupted transactions. In the view point of the service planning, this practice is atypical setup of taken offerings. The example in this paper is targeting 60K users, 1M USD total token values and the cost of backup nodes for the governance is 0.1 cents per a genuine node for the strategic alliance (see Table 2):

| Name | Value | Description |
|---|---|---|
| $N$ | 60,000 [User] | Total number of the nodes in the network |
| $V$ | 1,000,000 [USD] | (Target) total value of tokens (or coins) offered by ICO |
| $c_\eta$ | 0.1 [Cents/Member] | Cost for reserving nodes (i.e., alliance members) to avoid attacks |
| $\varrho$ | 0.7 | The acceptance rate for the alliance within members at $t_{\nu-1}$ |
| $C_0$ | 150 [Blocks] | Total number of blocks that changed by an attacker at $t_0(=0)$ |
| $J_k$ | 50 [Blocks] | Number of corrupted blocks by an attacker |

**Table 2.** Initial conditions for the cost function

It is noted that the values that described above are artificially given only for demonstration purposes. Since the *Strategic Alliance for BGG* model has been analytically solved, the values for the cost function and the probability distributions could be calculated straight forward (see Table 3).

| Name | Formula | Description |
|---|---|---|
| $q^0$ | (2.16), (3.7) | Probability that an attacker catches the blocks more than a half at $t_\nu$ |
| $q^1_\eta$ | (2.21), (3.7) | Probability that an attacker catches after adding reserved nodes |
| $\eta$ | $-$ | The number of alliance member for blockchain protection |
| $\eta^*$ | (3.4) | The optimal number of the strategic alliance nodes |
| $p_{c_{-1}}$ | (2.47), (3.8) | The probability that an attacker is not succeed at $t_{\nu-1}$ |
| $\mathfrak{S}(\eta)_{Total}$ | (3.6) | The total cost function for the enhanced blockchain network |

**Table 3.** Calculated values from the equations

However, it still requires the software implementation by using a programing language. Based on the conditions, the LP (Linear Programing) model could be described as follows:

*Objective* (3.9)

$$\min G = \mathfrak{S}(\eta)_{Total} \tag{4.41}$$

*Subject to* (3.4)

$$\eta \geq \frac{c_\eta}{V\cdot q^0 - c_\eta}; \tag{4.42}$$

$$\eta \le \tfrac{N}{2}; \tag{4.43}$$

From (3.9), the total cost $\mathfrak{S}(\eta)_{Total}$ is as follows:

$$\mathfrak{S}(\eta)_{Total} = \left(c_\eta\left(1-q_\eta^1\right) + (c_\eta+V)q_\eta^1\right)p_{c_{-1}} + V\cdot q^0(1-p_{c_{-1}}) \tag{4.44}$$

where

$$\begin{aligned} p_{c_{-1}} &= \boldsymbol{P}\left\{C_{\nu-1} < \tfrac{N}{2}\right\} \\ &\simeq \boldsymbol{P}\left\{C_\nu < \tfrac{N}{2} - \lambda_c\widetilde{\alpha}\ \right\} \\ &= \sum_{k=0}^{\{\frac{N}{2}-\lambda_c\widetilde{\alpha}\}} \left(\tfrac{\{\lambda_c(\widetilde{\alpha}_0+\mathbb{E}[\nu-1]\widetilde{\alpha})\}^k}{k!}\cdot e^{-\lambda_c(\widetilde{\alpha}_0+\mathbb{E}[\nu-1]\widetilde{\alpha})}\right), \end{aligned} \tag{4.45}$$

$$q^0 \simeq 1 - \sum_{k=0}^{\frac{N}{2}} \left(\tfrac{\{\lambda_c(\widetilde{\alpha}_0+\mathbb{E}[\nu-1]\widetilde{\alpha})\}^k}{k!}\cdot e^{-\lambda_c(\widetilde{\alpha}_0+\mathbb{E}[\nu-1]\widetilde{\alpha})}\right) \tag{4.46}$$

$$q_\eta^1 = \sum_{j=0}^{\eta} \sum_{\{k\ge\frac{N}{2}+j\}} \left(\tfrac{\lambda_c(\widetilde{\alpha}_0+\mathbb{E}[\nu-1]\widetilde{\alpha})}{k!}\cdot e^{-\lambda_c(\widetilde{\alpha}_0+\mathbb{E}[\nu-1]\widetilde{\alpha})}\right)P\{B_\eta = j\}, \tag{4.47}$$

$$P\{B_\eta = j\} = \binom{\eta}{j}\varrho^j(1-\varrho)^{\eta-j}. \tag{4.48}$$

As it is mentioned in Table 3, the parameter $\eta$ is the number of the strategic alliance members to protect the network from an attacker. The below illustration in Fig. 2 is atypical graph of an optimal result by using the *Strategic Alliance for Blockchain Governance Game* based on the given conditions (see Table 2). For this example, the optimal cost is 57.4K USD when the defender reserve the 7,000 alliance members with the 0.7 acceptance rate of the alliance action request.

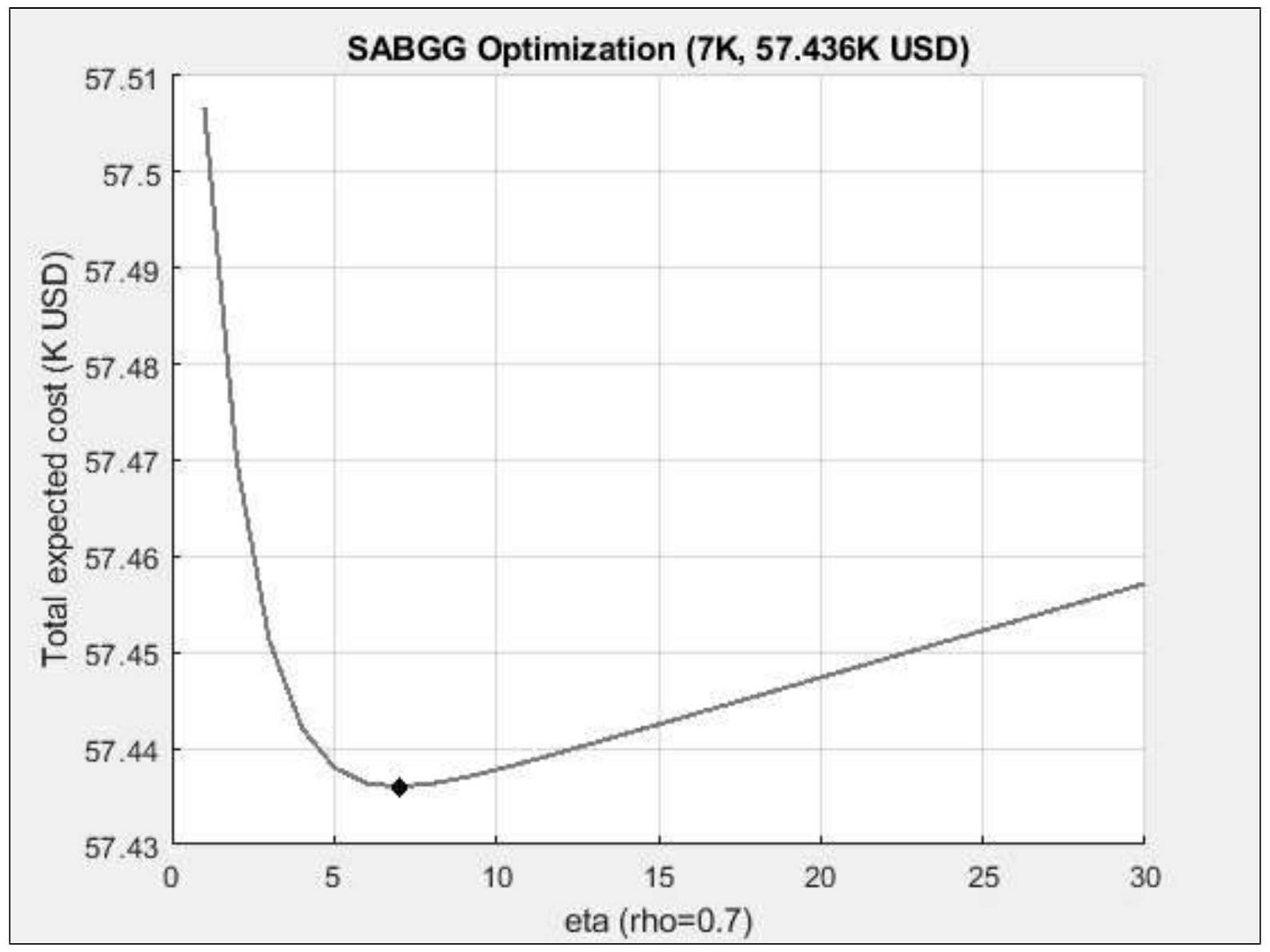

**Fig. 2.** Optimization Example for the Blockchain Governance Game

The moment of requesting the alliance of honest nodes will be the time $t_{\nu-1}$ when is one step prior than the time when an attacker catches more than half of the whole blocks (i.e., $t_\nu$).

## 5. CONCLUSION

This paper is establishing the enhanced theoretical framework of the *Blockchain Governance Game* by bring a new concept from the business area. This research includes the explicit equations for developing a new Blockchain network security to avoid the majority attack by doing the strategic alliance with other genuine miners. The main contributions of this research are the proof of the theorem of the *Strategic Alliance for BGG* (BGG-2) and the analytic functionals for the decision making parameters. In addition, the special case also fully analyzed for visualizing the results even though the setup of the case is artificially built. This research could be an alternative approach for whom considers the initial coin offering (ICO) or launching new Blockchain based services with enhancing the security features by adapting a strategic alliance. The SABGG is the first extended basic theoretical study from the BGG and this study could be extended further to various areas including cybersecurity, network architecture, service design and IT business model based on the Blockchain. In addition, further studies about setting up the initial parameters based on the real measured data could be yet another future research topics.